\documentclass[twocolumn,prc,aps,showpacs]{revtex4}
\usepackage{graphicx}

 \def\gsim{\mathrel{\rlap{\lower4pt\hbox{\hskip1pt$\sim$}}
 \raise1pt\hbox{$>$}}}

 \newcommand\la{\langle}
 \newcommand\ra{\rangle}
 \newcommand\beq{\begin{equation}}
 
 \newcommand\eeq{\end{equation}}
 \newcommand\beqn{\begin{eqnarray}}
 \newcommand\eeqn{\end{eqnarray}}
\def\mb{\,\mbox{mb}}
\def\fm{\,\mbox{fm}}
\def\GeV{\,\mbox{GeV}}

\def\lsim{\mathrel{\rlap{\lower4pt\hbox{\hskip1pt$\sim$}}
    \raise1pt\hbox{$<$}}}         
\def\gsim{\mathrel{\rlap{\lower4pt\hbox{\hskip1pt$\sim$}}
    \raise1pt\hbox{$>$}}}         
\def\J{J/\Psi}

\def\mb{\,\mbox{mb}}
\def\fm{\,\mbox{fm}}
\def\GeV{\,\mbox{GeV}}

\def\s0{\sigma_0(s)}

\def\beq{\begin{equation}}
\def\eeq{\end{equation}}

\def\beqy{\begin{eqnarray}}
\def\eeqy{\end{eqnarray}}

\newcommand{\ber}{\begin{displaymath}}
\newcommand{\eer}{\end{displaymath}}
\newcommand{\bey}{\begin{eqnarray}}
\newcommand{\eey}{\end{eqnarray}}

\def\beq{\begin{equation}}
\def\eeq{\end{equation}}

\def\beqy{\begin{eqnarray}}
\def\eeqy{\end{eqnarray}}

\begin{document}

\title{\bf  \boldmath$J/\Psi$ production in nuclear collisions: measuring the transport coefficient}

\vspace{1cm}

\author{B. Z. Kopeliovich}
\author{I. K. Potashnikova}
\author{Iv\'an Schmidt}
\affiliation{Departamento de F\'{\i}sica, Instituto de Avanzados en Ciencias en
Ingener\'{\i}a, Universidad T\'ecnica Federico Santa Mar\'{\i}a,
\\and\\
Centro Cient\'ifico-Tecnol\'ogico de Valpara\'iso,\\
Casilla 110-V, Valpara\'iso, Chile}
\begin{abstract}
\noindent The observed $p_T$-dependence of nuclear effects for $\J$
produced in heavy ion collisions at RHIC might look puzzling, since
the nuclear suppression seems to fade at large $p_T$. We explain
this by the interplay of three mechanisms: (i) attenuation of $\J$
in the hot medium created in the nuclear collision; (ii) initial
state shadowing of charmed quarks and attenuation of a $\bar cc$
dipole propagating through the colliding nuclei; (iii) a strong
Cronin effect for $\J$ caused by saturation of gluons in the
colliding nuclei. All three effects are well under control and
calculated in a parameter free way, except for the transport
coefficient $\hat q_0$ characterizing the medium. This is adjusted
to the $\J$ data and found to be in good agreement with the pQCD
prediction, but more than an order of magnitude smaller than what
was extracted from jet quenching data within the energy loss
scenario.

\end{abstract}



\maketitle

\section{Introduction}

\begin{figure}[htb]
\includegraphics[width=7cm]{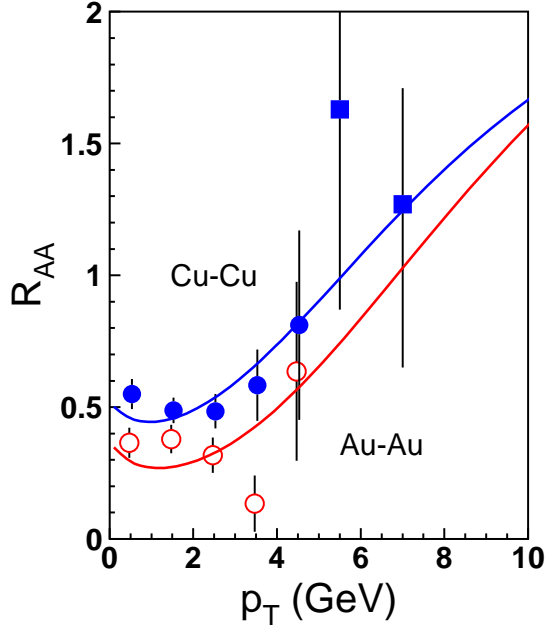}
\caption{\label{Cu+Au} (Color on line) Nuclear ratio $R_{AA}$ for central (0-20\%) copper-copper (full circles and squares, upper curve) and gold-gold (empty circles, bottom curve) as function of transverse momentum of the $\J$. The curves are calculated with Eq.~(\ref{500}) as is described in text.}
 \end{figure}

The recent measurements \cite{phenix1,phenix2,star1} of $\J$
produced in heavy ion collisions at RHIC have revealed unusual
features of the transverse momentum distribution. While all species
of hadrons measured so far demonstrate nuclear suppression, which
increases with $p_T$ and then levels off, the nuclei-to-$pp$ ratio
for $\J$ production, plotted in Fig.~\ref{Cu+Au} rises with $p_T$
and has even a tendency to exceed one \cite{star1}.

No explanation has been proposed so far, except for an exotic one
\cite{rapp} assuming that part of the production rate comes from
accidental coalescence of $c$ and $\bar c$ pairs available in the
medium. Even if this might happen, one should consider first of all
the conventional explanations, based on known dynamics.

We consider here three different mechanisms affecting the production
rate of $\J$ in heavy ion collisions: (i)~final state attenuation
(FSI) of $\J$ in the dense medium; (ii)~initial state interaction
(ISI), nuclear shadowing of charm quarks and the breakup of the
$\bar cc$ dipole propagating through the colliding nuclei; (iii)~ISI
Cronin effect for $\J$ caused by gluon saturation in the colliding
nuclei. All three effects certainly exist and are important, and
below we present their evaluation, which is performed in a parameter
free way, except the transport coefficient \cite{bdmps}
characterizing the hot medium. This is assumed to be unknown and is
adjusted to reproduce the data with the value of $\hat q_0\approx
0.3-0.5\GeV^2/\fm$. This value is an order or two of magnitude less
than what was extracted so far from high-$p_T$ pion suppression
observed in gold-gold collisions at $\sqrt{s}=200\GeV$, and
interpreted within the energy loss scenario \cite{d'enterria,urs}.

Only the first of the three effects mentioned above, the $\J$
attenuation due to FSI, was considered in the recent publication
\cite{wu-ma}, but the ISI suppression was ignored. Besides, the
$\bar cc$ separation was assumed to be fixed during propagation
through the medium, while the $\J$ wave function is fully formed
within a very short distance, half a fermi (see next section
\ref{evolution}). As a result,  $\hat q_0$ was grossly (5 times)
overestimated in \cite{wu-ma}.

On the contrary, in \cite{satz} it was assumed that $\J$ is
suppressed only by ISI, but propagates with no attenuation through
the produced dense matter. The observed nuclear effects were
explained by ISI and by the suppressed feed-down from the decays of
heavier states ($\chi,\Psi'$), which can be dissolved in the hot
medium. Such an approach does not look self-consistent: if $\J$ is
absorbed even in the cold nuclear matter, it should be even more
suppressed propagating through a dense medium.

\section{Final state interaction of \boldmath$\J$}
\subsection{ Time evolution of a small dipole}\label{evolution}

A $\bar cc$ dipole is produced at $x_F=0$ in the c.m. of the
collisions, with a short time scale $t^*_p\sim
1/\sqrt{4m_c^2+p_T^2}$ and with a small transverse separation $r\sim
1/m_c$. Then it evolves its size and forms the $\J$ wave function.
The full quantum-mechanical description of this process is based on
the path integral technique \cite{kz91}. However, just a rough
estimate of the formation time is sufficient here, since this time
scale turns out to be very short.

A small size dipole is expanding so fast that its initial size is
quickly forgotten. Indeed, the speed of expansion of a dipole
correlates with its size: the smaller the dipole is, the faster it
is evolving. This is controlled by the uncertainty principle, $k\sim
1/r$.
\beq \frac{dr}{dt}=\frac{2k}{E^*_c}\approx
\frac{4}{E^*_{\J}\,r}, \label{20}
\eeq
where $E^*_{\J}=2E^*_c$ is
the $J/\Psi$ energy in the c.m. of the collision; $k\sim 1/r$ is the
transverse momentum of the $c$-quark relative to the $\J$ direction.
The solution of this equation reads \beq
r^2(t)=\frac{8\,t}{E^*_{\J}} + r_0^2, \label{40} \eeq where
$r_0^2\sim 1/(p_T^2+m_c^2)$ is the initial dipole size squared,
which is neglected in what follows.

According to (\ref{40}) the expanding $\bar cc$ reaches the $\J$ size very fast,
\beq
t^*_f={1\over8}\,\la r_{\J}^2\ra\,\sqrt{p_T^2+m_{\J}^2}<0.6\fm,
\label{45}
\eeq
for $\J$ transverse momenta up to $5\GeV$. This is about the expected time of creation of the medium.

Another estimate of the formation time scale in the c.m. of collision is \cite{kz91},
\beq
t^*_f=\frac{2\,\sqrt{p_T^2+m_{\J}^2}}{m_{\Psi^*}^2-m_{\J}^2},
\label{50}
\eeq
where $m_{\Psi^*}$ is the mass of the first radial excitation.
This results in the same estimate as (\ref{45}).

We conclude that what is propagating through the medium is not a
small $\bar cc$ dipole (pre-hadron), but a fully formed $\J$.

\subsection{\boldmath$\J$  attenuation in a dense medium}

A charmonium propagates a path length $L$ in a medium with the
survival probability
\beq S(L)=\exp\left[-\int\limits_0^L dl\,
\sigma[r(l)]\,\rho(l)\right]. \label{60}
\eeq
Here the path length
and time are related as $l=vt$, with the $\J$ speed
$v=\sqrt{1-(2m_c/E)^2}$. The medium density is time dependent, and
is assumed to dilute as  $\rho(t)=\rho_0\,t_0/t$ due to the
longitudinal expansion.

The dipole cross section for small dipoles is $\sigma(r)=C\,r^2$,
where $r$ is the transverse $\bar cc$ separation. Correspondingly,
$\sigma_{\J}={2\over3}C\,\la r^2_{\J}\ra$, and we rely on the result
of the realistic model \cite{buchmuller,ihkt} for the mean $\J$
radius, $\sqrt{\la r^2_{\J}\ra}=0.42\fm$. The factor $C$ for
dipole-proton interactions is known from DIS data. Its value for a
hot medium is unknown, as well as the medium properties. However,
the factor C also controls broadening of a quark propagating through
the medium \cite{jkt,dhk},
\beq \Delta p_T^2(L) =
2\,\frac{d\sigma(r)}{dr^2}\Bigr|_{r=0} \int\limits_0^L dl\,\rho(l)
\label{80} \eeq
Thus, the factor $C$ is related to the transport
coefficient $\hat q$ \cite{bdmps}, which is in-medium broadening per
unit of length,
\beq C=\frac{\hat q}{2\,\rho}. \label{90} \eeq
So
one can represent the survival probability of $\J$ in the medium,
Eq.~(\ref{60}), as
\beq S(L)=\exp\left[-{1\over3}\,\la
r_{\J}^2\ra\int\limits_0^L dl\, \hat q(l)\right]. \label{100} \eeq

The transport coefficient depends on the medium density, which is a
function of impact parameter and time. We rely on the conventional
form \cite{frankfurt},
\beq \hat q(t,\vec b,\vec\tau)=\frac{\hat
q_0\,t_0}{t}\, \frac{n_{part}(\vec b,\vec\tau)}{n_{part}(0,0)},
\label{120} \eeq
where $\vec b$ and $\vec \tau$ are the impact
parameter of the collision and of the point where the $\hat q$ is
defined. The transport coefficient $\hat q_0$ corresponds to the
maximum medium density produced at impact parameter $\tau=0$ in
central collision ($b=0$) of two nuclei, at the time $t=t_0$ after
the collision. In what follows we treat the transport coefficient
$\hat q_0$ corresponding to the medium produced in central gold-gold
collision at $b=\tau=0$, as a adjusted parameter. It is rescaled for
other nuclei according to the number of participants $n_{part}(\vec
b,\vec\tau)$  \cite{frankfurt}. In what follows we consider
collision of identical nuclei, $A=B$, at $b=0$.

Eventually, integrating the attenuation factor Eq.~(\ref{100}) over
different direction of propagation of the $\J$ produced at impact
parameter $\vec\tau$ one gets  the FSI suppression factor in the
form, \beq R_{AA}^{FSI}(\vec\tau,p_T)\Bigr|_{b=0}= \int\limits_0^\pi
\frac{d\phi}{\pi} \exp\Biggl[-{1\over3}\la
r_{\J}^2\ra\int\limits_{l_0}^\infty dl \hat q(\vec\tau+\vec l)
\Biggr]. \label{140} \eeq
Here $|\vec\tau+\vec l|^2=\tau^2+l^2+2\tau
l\cos\phi$; and $l_0=vt_0$. The time scale $t_0$ for creation and
thermalization of the medium is rather uncertain, since gluons with
different transverse momenta are radiated at different coherence
times. We rely on the usual estimate $t_0=0.5\fm$.

\begin{figure}[htb]
\includegraphics[width=7cm]{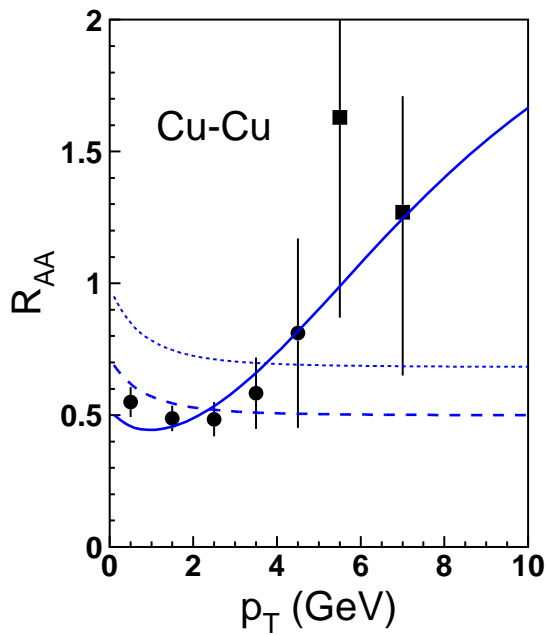}
\caption{\label{Cu} Data for the nuclear ratio $R_{AA}$ for central copper-copper collisions as function of transverse momentum of the $\J$ from \cite{phenix2} (circles) and \cite{star1} (squares).
The dotted curve shows the FSI effects,  Eq.~(\ref{140}).
The dashed curve includes the ISI effects, charm shadowing and absorption,  Eq.~(\ref{320}).
The solid curve is also corrected for the Cronin effect Eq.~(\ref{490}).
}
 \end{figure}

The results are depicted by dotted curve in Fig.~\ref{Cu} for copper-copper collisions.
We use $\hat q_0=0.45\GeV^2/\fm$, which allows to reproduce data well, provided that other corrections, discussed in the following sections, are added.

\section{Initial state Interaction: shadowing and absorption}

\subsection{Higher twist shadowing of charm}

The same time scales, production and formation, look very different
in the rest frame of one of the collision nuclei. While a fully
formed $\J$ propagates through the hot medium, in this reference
frame a $\bar cc$ dipole with a size "frozen" by the Lorentz time
dilation propagates through the cold nuclear matter. Both the
production and formation times become longer by the Lorentz factor
$\gamma=2E_{\Psi}/\sqrt{4m_{c}^2+p_T^2}$. The coherence time of
$\bar cc$ pair production reads, \beq
t_c=\frac{E_{\J}}{(4m_c^2+p_T^2)} =
\frac{2m_{c}\,\sqrt{s}}{(4m^2_{c}+p_T^2)\,m_N}, \label{160} \eeq
and
is rather long. At $\sqrt{s}=200\GeV$ it varies from $13$ to $4\fm$
for $0<p_T<5\GeV$, i.e. is of the order of the nucleus size, or
longer. Correspondingly, the formation time is even longer,
$t_f\approx 5t_c\gg R_A$.

A full calculation of the nuclear effects for $\J$ produced in $pA$
collisions, including the effects of shadowing and breakup
interactions of the final $\bar cc$, has not been done so far. Only
production of the $\chi$, the P-wave charmonium, which is a simpler
case, was calculated in detail in \cite{kth-psi}. Besides, a
considerable fraction of $\J$s are produced via decay of heavier
states, $\Psi'$, $\chi$, etc. For our purposes it would be safer to
use the experimental value of nuclear suppression observed in $d-Au$
collisions at $\sqrt{s}=200\GeV$ \cite{phenix-dA}. Unfortunately the
experimental uncertainty is still large, so we fix
$R_{dA}(x_F=0)=0.8$ which is about the central value.

Even if this nuclear suppression factor integrated over impact
parameter is known, it is not sufficient to perform calculations for
$AA$ collisions. One has to know the $b$-dependence of $R_A$. Since
no relevant data are available so far, we can only rely on the
theory, being constrained by the integrated value of $R_A$.

Since the coherence time Eq.~(\ref{160}) in the rest frame of one of
the colliding nuclei is rather long, we assume that the $\bar cc$
transverse separation is "frozen" by Lorentz time delation. This
grossly simplifies the calculations. In the dipole approach the
nuclear suppression factor caused by initial state $c$-quark
shadowing and attenuation of the $\bar cc$ dipole, has the form
\cite{kth-psi},

\begin{widetext}
\beqn
R_{NA}(\tau)&=&
{1\over T_A(\tau)}\left[\sum\limits_{\lambda=\pm}
\Biggl|\int d^2r\, \Psi^*_{\chi}(r)
\left(\vec e_{\lambda}\cdot\vec r\right)\Psi_{g\to\bar cc}(r)\Biggr|^2\right]^{-1}
\int\limits_{-\infty}^\infty dz\,\rho_A(\tau,z)\,
\sum\limits_{\lambda=\pm}
\Biggl|\int d^2r\, \Psi^*_{\chi}(r)
\left(\vec e_\lambda\cdot\vec r\right)\Psi_{g\to\bar cc}(r)
\nonumber\\ &\times&
\exp\left\{-{1\over2}\,\sigma_8(r)\,T_A(\tau,z)-
{1\over2}\,\sigma_{dip}(r)\left[T_A(\tau)-T_A(\tau,z)\right]\right\}\Biggr|^2
\label{200}
\eeqn
\end{widetext}
Here $T_A(\tau,z)=\int_{-\infty}^z dz'\,\rho_A(\tau,z')$; $\vec
e_\pm=(\vec e_x\pm\vec e_y)/\sqrt{2}$ is the polarization vector of
the projectile gluon. The light-cone wave function of a $\bar cc$ in
the gluon $\Psi_{g\to\bar cc}(r)\propto K_0(m_c r)$ \cite{kth-psi},
where $K_0$ is the modified Bessel function. Thus, the mean
transverse size squared of a $\bar cc$ fluctuation of a gluon, $\la
r^2\ra =1/m_c^2$ is small, at least an order of magnitude  smaller
than that of charmonia. Apparently, in a convolution of two
$r$-distributions, narrow and wide, the mean size is controlled by
the narrow one.

The cross section $\sigma_{dip}(r)$ in (\ref{200}) is the universal
dipole-proton cross section \cite{zkl}, which we use in the
small-$r$ approximation $\sigma_{dip}(r)\approx C(x_2)\,r^2$. The
factor $C(x_2)$ is calculated in \cite{broad} as function of $x_2$,
which is the fractional light-cone momentum of the $\J$ relative to
the target.

Shadowing of the process $g\to\bar cc$ is controlled by the
three-body $g$-$q$-$\bar q$ dipole cross section, which can be
expressed via the conventional dipole cross sections,
\beq
\sigma_8(r)={9\over8}\left[\sigma_{dip}(r_1)+\sigma_{dip}(r_2)\right]-
{1\over8}\sigma_{dip}(\vec r_1-\vec r_2), \label{220} \eeq
where
$\vec r_1$ and $\vec r_2$ are the transverse vectors between the
gluon and the $q$ and $\bar q$ respectively. We neglect the
distribution of the fractional light-cone momentum of the $q$ and
$\bar q$, fixing it at equal shares. Then $\vec r_1=-\vec r_2=\vec
r/2$, so that \beq \sigma_8(r)={7\over16}\,\sigma_{dip}(r).
\label{230} \eeq Here we also rely on the small-$r$ approximation.

Since we fixed the overall suppression $R_{dA}$ at the measured
value, and  need to know only the impact parameter dependence, a
rough estimate of Eq.~(\ref{200}) should be sufficient. Therefore,
we approximate the result of integration over $r$ in (\ref{200})
replacing the dipole cross sections by an effective cross section,
$\sigma_{dip}(r)\Rightarrow\sigma_{eff}$, which we can adjust to
reproduce $R_{dA}$. Then the suppression factor Eq.~(\ref{200})
takes the form,
\beq
R_{NA}(\tau)=\frac{16}{9\sigma_{eff}T_A(\tau)}\left[e^{-{7\over16}\sigma_{eff}T_A(\tau)}-
e^{-\sigma_{eff}T_A(\tau)}\right]. \label{240} \eeq
The first term
in square brackets represents shadowing, the second one is related
to the survival probability of the produced colorless $\bar cc$
dipole.

We extracted the value of $\sigma_{eff}$ comparing the integrated suppression,
\beq
R_{NA}={1\over A}\int d^2\tau\,T_A(\tau)\,R_{NA}(\tau),
\label{260}
\eeq
with data \cite{phenix-dA} for deuteron-gold collisions at $\sqrt{s}=200\GeV$, $x_F=0$. We fixed the measured ratio at $R_{dAu}=0.8$, and found $\sigma_{eff}=2.3\mb$.

This value can be compared with the theoretical expectation. As was
mentioned, in the convolution of the narrow distribution
$\Psi_{g\to\bar cc}(r)$ with the large size charmonium wave
function, the latter can be fixed at $r=0$, and the mean separation
is fully controlled by the $\bar cc$ distribution in a gluon. Then
the mean separation squared of a produced $\bar cc$ pair, i.e. a
fluctuation which took part in the interaction, is given by
\beq
\la r^2\ra = \frac{\int d^2r\,r^4\,K_0^2(m_cr)}{\int d^2r\,r^2\,K_0^2(m_cr)}=
{16\over5\,m_c^2}.
\label{280}
\eeq

Now we are in a position to evaluate the effective cross section,
\beq \sigma_{eff}=C(E)\,\la r^2\ra. \label{300} \eeq The energy
dependent factor $C(E)$ is calculated in \cite{broad}. At the energy
of $\J$ $E=300\GeV$ ($x_F=0,\ \sqrt{s}=200\GeV$) this factor varies
between $C=4.5$ in the leading order, down to $C=3.5$ if higher
order corrections are included. Correspondingly, the effective cross
section Eq.~(\ref{300}) range is $2.5\mb>\sigma_{eff}>2\mb$, which
is in excellent agreement with the value extracted from the RHIC
data.

We calculate the ISI suppression nucleus-nucleus collisions assuming
that the suppression factors due to simultaneous propagation of the
$\bar cc$ pair through both nuclei factorize. We ignore the possible
dynamics which can breakdown this assumption \cite{hk,hhk,hkp}, so
that the ISI suppression factor for a collision of nuclei $A$ and
$B$ with impact parameter $b$ reads,
\beq R_{AB}^{ISI}(\vec b)=\frac
{\int d^2\tau\, T_A(\tau)T_B(\vec b-\vec\tau)
R_{NA}(\vec\tau)R_{NB}(\vec b-\vec\tau)}
{\int d^2\tau\, T_A(\tau)T_B(\vec b-\vec\tau)}.
\label{320}
\eeq

Thus, the initial state interactions cause the additional
suppression, Eq.~(\ref{320}), of $\J$ produced in heavy ion
collision. The combined effect of ISI and FSI suppression in
copper-copper central collision is shown by the dashed curve in
Fig.~\ref{Cu}. While it agrees with the data at $p_T<3\GeV$, there
is indication that data at higher $p_T$ are underestimated.

\subsection{Leading twist gluon shadowing}

Besides quark shadowing, which is a higher twist effect and scales
as $1/m_c^2$, the leading twist gluon shadowing, which depends on
$m_c$ logarithmically,  may be important, depending on kinematics.
In terms of the Fock state decomposition, gluon shadowing is related
to higher Fock components in the projectile gluon, e.g. $g\to \bar
qqg$. Even this lowest state is heavier than just a $\bar qq$ and
should have a shorter coherence time. In terms of Bjorken $x$ this
means that the onset of gluon shadowing is shifted towards low $x_2$
compared to quark shadowing. Indeed, calculations \cite{kst2} show
that no gluon shadowing is possible above $x_2\approx 10^{-2}$.
Moreover, it was found in \cite{krt} that the coherence length which
controls the onset of gluon shadowing, is scale independent, i.e. it
the same for light and heavy quarks. This result of \cite{krt} can
be understood via the energy denominator for the $g\to\bar qqg$
transition amplitude,
\beq A(g\to\bar qqg)\propto \frac{1}{k^2+\alpha_g M_{\bar qq}^2},
\label{340} \eeq
where $k$ and $\alpha_g$ are the transverse and fractional
light-cone momenta of the radiated gluon, respectively. The factor
$\alpha_g$, which is predominantly small, suppresses the mass term
in (\ref{340}). Besides, the mean transverse momentum of gluons was
found in \cite{kst2} to be rather large, $\sqrt{\la
k^2\ra}=0.7\GeV$. This is dictated by data on large mass diffraction,
which is strongly suppressed compared to usual pQCD expectations.
This phenomenon has been known in the Regge phenomenology as
smallness of the triple Pomeron vertex. The large value of
$\sqrt{\la k^2\ra}$, also supported by many other experimental
evidences \cite{spots}, leads to suppression of the gluon radiation
amplitude Eq.~(\ref{340}) and weak gluon shadowing \cite{cronin}.
The latter is confirmed by a NLO analysis of the DIS data
\cite{florian}, but contradicts the recent analysis of \cite{eps08},
which resulted in a very strong gluon shadowing breaking the
unitarity bound \cite{bound}.

The coherence length available for gluon shadowing can also be
related to the value of $x_2$ \cite{krt},
\beq
l_c^g=\frac{P^g(1-x_1)}{x_2\,m_N}. \label{360} \eeq
Here the
denominator presents the usual Ioffe time scale, which is as long as
$(0.2\fm)/x_2$, and may exceed the nuclear size at small $x_2$. The
factor $1-x_1$ is usually neglected, assuming that $x_1$ is small,
which is not always the case. More important is the factor
$P^g\approx 0.1$ evaluated in \cite{krt}. Its smallness is actually
due to the large intrinsic transverse momentum of gluons, which we
have just discussed above.

Usually $x_2$ is defined as
\beq
x_2=e^{-\eta}\,\sqrt{\frac{m_{\J}^2+p_T^2}{s}}.
\label{380}
\eeq
It varies with pseudorapidity and reaches a minimum at the largest
measured value of $\eta$. At $\eta=0$, with the mean value of $\la
p_T^2\ra=4\GeV^2$ \cite{phenix-dA}, one gets $x_2=0.02$, which is
certainly too large for gluon shadowing. Therefore we can safely
disregard this correction in further calculations, done at $\eta=0$.

Notice that of course $x_2$ decreases with $\eta$ and reaches its
minimal value $x_2=2.5\times10^{-3}$ at the maximal rapidity
$\eta=2$. Although this value of $x_2$  allows some amount of gluon
shadowing, we expect a tiny correction. Indeed, within the color
singlet model (CSM) \cite{csm} and its modified version
\cite{ryskin}, which provides so far the only successful description
of $\J$ production in $pp$ collisions, the actual $x_2$ is
considerably larger than the value given by the usual definition
Eq.~(\ref{380}). This is because in the CSM $\J$ is produced
accompanied by a gluon, and their total invariant mass $M_{g\J}$ is
considerably larger than $m_{\J}$. With the mass distribution,
$d\sigma/dM_{g\J}^2\propto M_{g\J}^{-6}$, one gets the mean
invariant mass squared,
\beq
\left\la M_{g\J}^2\right\ra=2\,m_{\J}^2,
\label{400}
\eeq
which leads to a new more correct value $\tilde x_2\approx2x_2$.
With the corrected minimal value $\tilde x_2(\eta=2)=0.005$ gluon
shadowing correction is tiny, just a few percent \cite{kst2,krtj}.

\section{Broadening of gluons, Cronin effect}

In $pA$ collisions projectile gluons propagating through the nucleus
experience transverse momentum broadening due to multiple
collisions. As a result, the mean transverse momentum of produced
charmonia is larger than in $pp$ collisions. The dipole approach
\cite{jkt,dhk} is rather successful predicting broadening for heavy
quarkonia \cite{broad} and heavy Drell-Yan dileptons
\cite{e866-our}, $\Delta{pA}=\la p_T^2\ra_{pA}-\la p_T^2\ra_{pp}$ in
a parameter free way, relying on the phenomenological cross section
\cite{kst2} fitted to photoproduction and DIS data. Broadening for a
gluon of energy $E$ propagating a nuclear thickness $T_A$ reads
\cite{broad},
\beqn
\Delta_{pA}(E)=
{9\over16}\,\la T_A\ra\,\sigma^{\pi p}_{tot}(E)\,\left[Q_{qN}^2(E)
+\frac{3}{2\, \la r_{ch}^2\ra_\pi}\right],
\label{420}
\eeqn
where the proton saturation scale is
\beq
Q_{qN}(E) = 0.19\GeV\times\left(\frac{E}{1GeV}\right)^{0.14}.
\label{440}\\
\eeq

In fact, the broadening Eq.~(\ref{420}) is the saturation scale in
the nucleus calculated in the leading order, i.e. without
corrections for gluon saturation in the medium. Those corrections
lead to about $20\%$ reduction of $\Delta_q$ \cite{broad}.

Remarkably, broadening does not alter the shape of the
$p_T$-distribution of produced $J/\Psi$. Indeed,  data on  $pp$,
$pA$ and even $AA$ collisions, at the energies of fixed target
experiments \cite{fixed} and at RHIC \cite{phenix-dA}, are described
well by the simple parametrization,
\beq
\frac{d\sigma}{dp_T^2} \propto
\left(1+\frac{p_T^2}{6\la p_T^2\ra}\right)^{-6},
\label{460}
\eeq
where $\la p_T^2\ra$ is the mean transverse momentum squared, which varies dependent on the process.
Therefore, the simplest way to calculate the $p_T$-dependence of the nuclear cross section would be just making a shift $\Delta$  in the mean value $\la p_T^2\ra$ for $pA$ compared to $pp$, where $\Delta$ is broadening given by Eq.~(\ref{420}). Then the nuclear ratio as function of $p_T$ gets reads,
\beqn
R_{pA}(p_T)&=&\frac{\la p_T^2\ra\,R_{pA}}{\la p_T^2\ra+\Delta_{pA}}\,
\left(1+\frac{p_T^2}{6\la p_T^2\ra}\right)^6
\nonumber\\ &\times&
\left(1+\frac{p_T^2}{6[\la p_T^2\ra+\Delta_{pA}]}\right)^{-6},
\label{480}
\eeqn
where $R_{pA}$ is the $pA$ over $pp$ ratio of $p_T$-integrated cross sections, Eq.~(\ref{260}).

\begin{figure}[htb]
\includegraphics[width=6cm]{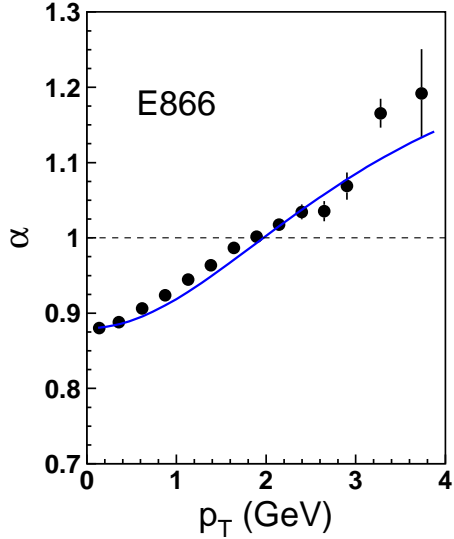}
\caption{\label{e866} The exponent $\alpha=1+\ln(R_{pA})/\ln(A)$ as function of $p_T$ calculated with Eq.~(\ref{480}) in comparison with data from the E866 experiment \cite{e866}.}
 \end{figure}

This simple procedure looks natural, although is not really proven.  We can test it with the precise data from the E866 experiment at Fermilab at $\sqrt{s}=39\GeV$. All the input parameters in (\ref{480}) are known from the same measurements \cite{e866,e866-pt} and other experiments at the same energy \cite{e772,e789,mmp} and also from our calculations, Eq.~(\ref{420}).
$\la p_T^2\ra=1.5\GeV^2$; $\Delta=0.08\GeV^2\times A^{1/3}$; $R_{pA}=A^{-0.05}$.
The $A$-dependence of the nuclear ratio calculated with Eq.~(\ref{480}) as function of $p_T$ is compared with E866 data in Fig.~\ref{e866} for the exponent characterizing the $A$-dependence, $\alpha=1+\ln(R_{pA})/\ln(A)$.
This comparison confirms the validity of the chosen procedure, at least within the measure interval of $p_T$.

Inspired by the good agreement with $pA$ data, we apply the same procedure to evaluation of the Cronin effect  for $\J$ production in central $AA$ collisions.
One can use for $R_{AA}(p_T)$ the same Eq.~(\ref{480}), which should be modified replacing $\Delta$ and $\la p_T^2\ra$ by corresponding values for $AA$ collisions at $\sqrt{s}=200\GeV$. Also one should replace the ratio of the $p_T$-integrated cross sections $R_{pA}\Rightarrow R_{AA}^{ISI}(b=0)$, which is calculated with Eq.~(\ref{320}).
\beqn
R_{AA}^{ISI}(b=0,\tau,p_T)&=&\frac{\la p_T^2\ra\,R_{AA}^{ISI}(b=0,\tau)}{\la p_T^2\ra+\Delta_{AA}(\tau)}\,
\left(1+\frac{p_T^2}{6\la p_T^2\ra}\right)^6
\nonumber\\ &\times&
\left(1+\frac{p_T^2}{6[\la p_T^2\ra+\Delta_{AA}(\tau)]}\right)^{-6}\!\!\!\!,
\label{490}
\eeqn
where the ratio of $p_T$-integrated cross sections, $R_{AA}^{ISI}(b=0,\tau)$, is given by Eq.~(\ref{320}) without integration over $\tau$.

According to Eqs.~(\ref{420})-(\ref{440}) broadening slowly rises with energy. However, the $\J$ energy in the nuclear rest frame is about the same in the E866 experiment ($\la E\ra=230\GeV$) and in Phenix data at $x_F=0$ and $\sqrt{s}=200\GeV$ ($E=330\GeV$), so we neglect the difference. Thus, in $AA$ collisions $\Delta $ simply doubles compared to $pA$, and we get $\Delta=0.64\GeV^2$ for coper-coper and $\Delta=0.93\GeV^2$ for gold-gold collisions.
Notice that the mean value of transverse momentum squared in $pp$ collisions at $\sqrt{s}=200\GeV$, $\la p_T^2\ra=4\GeV^2$, is considerably larger than at $\sqrt{s}=39\GeV$.

Eventually, we are in a position to combine all three effects and calculate the $p_T$-dependence of the nuclear ratio in central $AA$ collisions,

\beq
R_{AA}^{\J}(b=0,p_T)=\frac{\int\limits_0^\infty d^2\tau\,T_A^2(\tau)\,R_{AA}^{FSI}(\tau,p_T)\,
R_{AA}^{ISI}(\tau,p_T)}
{\int\limits_0^\infty d^2\tau\,T_A^2(\tau)}.
\label{500}
\eeq
The result is depicted by solid curve in Fig.~\ref{Cu} in comparison with data for copper-copper collision.
Calculations and data for gold-gold collisions are also shown in Fig.~\ref{Cu+Au}.

\begin{figure}[htb]
\includegraphics[width=7cm]{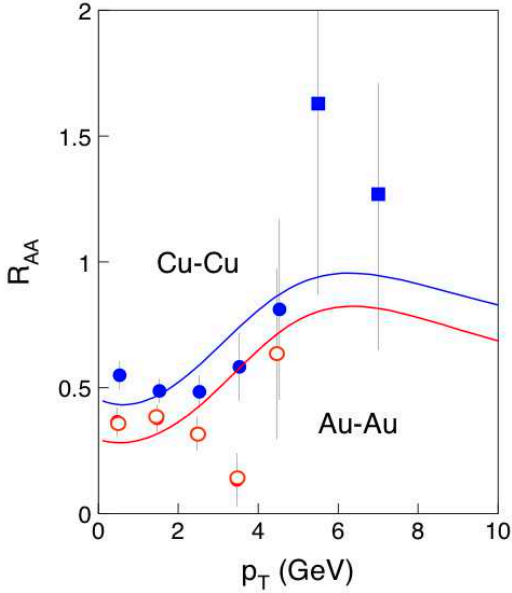}
\caption{\label{weak-cronin} The same as in Fig.~\ref{Cu+Au}, but the Cronin effect is calculated differently, with Eq.~(\ref{520}).}
 \end{figure}

Notice that the procedure  Eq.~(\ref{480}) has not been confronted with data above $p_T=4\GeV$,
so our extrapolation and predicted steep rise of the ratio at $p_T$, which might create problems with $k_T$-factorization, is not well justified. For this reason we tried another way to implement broadening into the $p_T$-distribution.
An alternative way would be a simple convolution of broadening, which we take in the Gaussian form, with the $p_T$ distribution in $pp$ collisions.
Then the nuclear modification factor $R_{AA}^{ISI}(\tau,p_T)$ get the form,
\beqn
R_{AA}^{ISI}(\tau,p_T)&=&\frac{R_{AA}^{ISI}}{\pi\Delta(\tau)}\left(1+\frac{p_T^2}{6\la p_T^2\ra}\right)^{6}
\int d^2k\,e^{-k^2/\Delta(\tau)}
\nonumber\\ &\times&
\left(1+\frac{(\vec p_T-\vec k)^2}{6\la p_T^2\ra}\right)^{-6}
\label{520}
\eeqn
The results for copper-copper and lead-lead collisions are plotted in Fig.~\ref{weak-cronin}.
We see that the description of data at $p_T<5\GeV$ is unchanged compared to what was depicted in Fig.~\ref{Cu+Au}. This means that
our determination of $\hat q_0$ from the data is stable against the choice of the way how the $p_T$ broadening is included. Only the behavior at larger $p_T$, where available data have poor accuracy, is altered, showing a weaker Cronin enhancement.

\section{Probing dense matter at SPS}

\begin{figure}[htb]
\includegraphics[width=6cm]{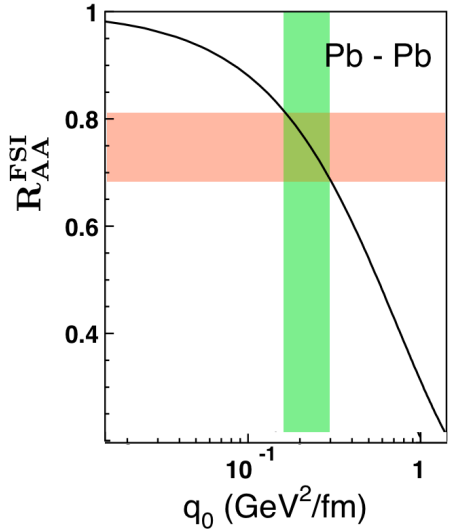}
\caption{\label{sps-fsi} The FSI attenuation factor for $\J$ produced in central lead-lead collisions at SPS.
The curve corresponds to suppression versus $\hat q_0$ calculated with Eq.~(\ref{500} with excluded ISI factor and integrated over $p_T$. The horizontal stripe shows the magnitude and uncertainty of suppression reported by the NA60 experiment \cite{na60}. The vertical stripe shows the interval of values of $\hat q_0$ which allow to describe the observed suppression.}
 \end{figure}

The nuclear suppression caused by FSI of the $J/\Psi$ with the produced medium was determined in the NA50 and NA60 experiments comparing the measured nuclear suppression $R_{AA}$ with what one could expect as the cold nuclear effects in initial state extrapolating from $pA$ data. The latest results from the NA60 experiment  \cite{na60} for maximal number of participants corresponding to central collisions show that the FSI suppression factor is $0.75\pm0.7$. This experimental uncertainty is shown by the horizontal stripe in Fig.~\ref{sps-fsi}.
The curve shows dependence of the FSI suppression factor $R_{AA}^{FSI}(b=0)$, calculated with Eq.~(\ref{500}), on the transport coefficient. The factor $R_{AA}^{ISI}$ was excluded from (\ref{500}) and integration over $p_T$ performed.
Together with the experimental uncertainty this curve provides an interval of values of $\hat q_0=0.23\pm0.07$ (vertical stripe), which allows to describe the observed "anomalous" suppression. It is about twice as small as we got from RHIC data.

\section{Summary and outlook}

We performed an analysis of data for $p_T$-dependent nuclear effects
in $\J$ production in central copper-copper and gold-gold collisions
observed at RHIC. These data, looking puzzling at first glance, have
not received a proper interpretation so far.  We evaluated the final
state attenuation of the produced $\J$ in the created dense medium
relating it to the transport coefficient, i.e. broadening of partons
propagating through the medium. The key point, which allows to
establish this relation, is the dipole description of broadening
\cite{dhk,jkt}.

The observed nuclear effects in $\J$ production in $AA$ collisions
is interpreted as a combination of FSI of the fully formed $\J$ in
the dense medium, and the ISI effects in production of $\J$ caused
by multiple interactions of the colliding nuclei. The latter
includes, attenuation of the produced $\bar cc$ dipole propagating
through both nuclei, higher twist shadowing of charm quarks, and
leading twist gluon shadowing. Besides, gluon saturation in nuclei
leads to a considerable broadening of gluons, which causes a strong
Cronin effect for $\J$. This explains the observed  unusual rise of
the nuclear ratio with $p_T$.

All effects are evaluated in a parameter free way, except for the
unknown properties of the produced hot medium. We employed the
conventional model for the space-time development of the produced
matter relating it to the number of participants. The only parameter
adjusted to data, $\hat q_0$, is the transport coefficient
corresponding to a maximal density of the matter produced in central
gold-gold collision. We found that the $\J$ data from RHIC are well
reproduced with $\hat q_0\approx 0.3-0.5\GeV^2/\fm$. This is close
to the expected value $\hat q_0=0.5\GeV^2/\fm$ \cite{bdmps}, and
more than order of magnitude less than was found from jet quenching
data within the energy loss scenario \cite{phenix-theory}.

We also examined the $\J$ data from the NA60 experiments at SPS,
which are available for $p_T$-integrated cross sections, and with
already separated ISI effects. From the observed suppression in
central lead-lead collisions we found $\hat q_0\approx
0.23\pm0.07\GeV^2/\fm$.

We performed the calculations assuming direct $\J$ production, but
it is known that about $40\%$ comes from the feed-down by decays of
heavier charmonium states, $\chi$ and $\Psi'$ \cite{feed-down}.
Those states are about twice as big as the $\J$ \cite{ihkt}, and
correspondingly should have a larger absorption cross section.
Therefore, adding these channels of $\J$ production can only reduce
the value of $\hat q_0$, i.e. the above values should be treated as
an upper bound. The bottom bound can be estimated assuming that the
$r^2$-approximation is valid up to the size of these excitations
(which is certainly an exaggeration). Then the bottom bound for
$\hat q_0$ is the above values times a factor $0.7$.

We conclude that $p_T$-dependent nuclear effects for $\J$ production
in heavy ion collisions provide a sensitive probe for the dense
medium produced in these collisions. Both experimental and
theoretical developments need further progress. More accurate $pA$
(or $dA$) data are needed for a better control of the ISI effects in
$AA$ collisions. Data for other heavy quarkonia ($\Psi',\chi,
\Upsilon$) would bring new precious information.

On the theoretical side, more calculations are required to describe
the observed centrality and rapidity dependence of nuclear effects.
The azimuthal asymmetry can also be calculated. The small-$r$
approximation for the dipole cross section used here may not be
sufficiently accurate for a very dense matter. One should rely on a
more elaborated $r$-dependence. A more realistic model for the
space-time development of the dense medium, including transverse
expansion, should be considered.

\vspace{3mm}

\begin{acknowledgments}

We are grateful to Dima Kharzeev for informative  discussions.
This work was supported in part by Fondecyt (Chile) grants 1090236,
1090291 and 1100287,  by DFG (Germany) grant PI182/3-1, and by
Conicyt-DFG grant No. 084-2009.

\end{acknowledgments}

\end{document}